\documentclass[a4paper,aps,pre,floatfix,twocolumn, superscriptaddress,nobibnotes,nolongbibliography,nofootinbib]{revtex4-2}

\usepackage[pdftex]{graphicx}
\setkeys{Gin}{width=0.9\columnwidth}

\usepackage{times}
\usepackage{verbatim}
\usepackage{bbold}
\usepackage{bbm}
\usepackage{bm}
\usepackage{lipsum}

\usepackage{amsmath}
\usepackage{amsfonts}
\usepackage{color}
\usepackage[english]{babel}
\usepackage{microtype}
\usepackage{comment}
\usepackage[normalem]{ulem}
\usepackage{float}

\usepackage{soul,xcolor}
\setstcolor{red}

\usepackage{fancyhdr}
\usepackage{relsize}

\usepackage[hidelinks,unicode=true]{hyperref}
\hypersetup{
colorlinks=true,
linkcolor=blue,
urlcolor=blue,
citecolor=blue,
pdfhighlight=/N}

\usepackage{comment}
\usepackage{subcaption}

\newcommand{\av}[1]{\langle #1 \rangle}

\begin{document}


\title{Modeling financial transactions via random walks on temporal networks}

\author{Carolina E. Mattsson}
\email{Corresponding author: mattsson.c@northeastern.edu}
\affiliation{Innovation \& BSR, Intesa Sanpaolo, 10138 Torino, Italy}

\author{Claudio Cellerini}
\affiliation{Department of Physics, Universit\'a di Torino, 10125 Torino, Italy}

\author{Jaume Ojer}
\affiliation{Institut de F\'isica Interdisciplin\`aria i Sistemes Complexos (IFISC, CSIC-UIB), 07122 Palma de Mallorca, Spain}

\author{Michele Starnini}
\email{Corresponding author: michele.starnini@upf.edu}
\affiliation{Department of Engineering, Universitat Pompeu Fabra, 08018 Barcelona, Spain}

\date{\today}

\begin{abstract}
We model financial transactions as random walks on activity-driven temporal networks. 
By enforcing fund conservation, 
our framework analytically derives heavy-tailed distributions for the
stationary balances and 
transaction sizes. 
Crucially, the latter is driven by variance 
in the spending propensity of individuals. 
Calibrated with empirical data from a closed, digital currency community, the model also reproduces observed correlations between inflows and outflows. 
Our findings provide a path for understanding 
emergent properties of the circulation of money.
\end{abstract}

\maketitle

\section{Introduction}

Economies are the complex systems that sustain human civilization, coordinating production, trade, and consumption across billions of interacting agents~\cite{arthur_foundations_2021}. 
The emergent properties of these systems have long attracted attention, from world trade networks~\cite{garlaschelli_structure_2005} to growth~\cite{mcnerney_how_2022, mattsson_network_2025} and systemic risk~\cite{battiston_debtrank_2012, diem_quantifying_2022}. Universalities in firm-size distributions~\cite{axtell_zipf_2001}, wealth concentration~\cite{mandelbrot_pareto-levy_1960, bouchaud_wealth_2000}, and network structure~\cite{boss_network_2004, bacilieri_firm-level_2023} hint at underlying mechanics that transcend institutional particulars. Indeed, the coordinating role of exchange across economic systems has led the physics community to explore ``asset-exchange models''~\cite{ispolatov_wealth_1998, patriarca_basic_2010, greenberg_twenty-five_2024}. The dynamics where populations of agents transfer assets among themselves under simple exchange mechanisms have been linked to kinetic energy exchange among gas molecules~\cite{chakraborti_statistical_2000, dragulescu_statistical_2000} and to specific interpretations of economic exchange~\cite{chakrabarti_microeconomics_2009}. Despite the theoretical appeal of these models, they face significant limitations, most notably in calibration and validation against real economic systems~\cite{gallegati_worrying_2006}.

Payments---commonly used to settle exchanges of goods, services, or assets---offer a way forward. Modern digital payment systems produce exhaustive records, and these proprietary datasets are becoming available for research~\cite{letizia_corporate_2019, semeraro_structural_2020, starnini_smurf-based_2021, mattsson_trajectories_2021, fujiwara_money_2021, ialongo_reconstructing_2022} and national statistics~\cite{bacilieri_firm-level_2023, hotte_mapping_2025}. Cryptocurrencies make transactions public by design, albeit distorted by privacy-preserving tactics~\cite{meiklejohn_fistful_2016, coquide_orbitaal_2024}. The newly available datasets have surfaced several empirical regularities regarding the circulation of money. At the granularity of individual firms, total payment inflows and outflows scale one-to-one~\cite{bacilieri_firm-level_2023}, consistent with a locally conservative process. Moreover, payment dynamics are highly heterogeneous~\cite{campajola_microvelocity_2022, mattsson_inverse_2023, collibus_microvelocity_2025} and transaction-size distributions are heavy-tailed, spanning many orders of magnitude~\cite{starnini_smurf-based_2021}.

A fundamental physical constraint in payment systems is the conservation of funds: One must have money to spend money. Checks bounce and digital payments are declined when transactions cannot be covered, blockchain protocols are engineered to prevent double-spending, and cash systems rely on steep punishments for counterfeiting. Glitches in this conservation law are rare enough to be newsworthy~\cite{rivlin_why_2015, partridge_bank_2023}. Systematic violations arise with issuance or dissolution and mark the boundaries of open payment systems~\cite{mattsson_trajectories_2021}.

In this Letter, we propose a simple yet flexible model that characterizes financial transactions as simultaneous random walks on an activity-driven temporal network~\cite{perra_activity_2012, starnini_topological_2013}. 
We validate our model using a comprehensive, pseudonymized dataset from Sarafu, a digital community currency in Kenya~\cite{ruddick_sarafu_2021, mattsson_sarafu_2022}. 
We derive an analytical formulation for the stationary distribution of funds across the population, i.e., the balance distribution. 
We analytically demonstrate that a power-law decay with exponent $-2$ in the balance distribution, reported also for previous models~\cite{chatterjee_kinetic_2007}, is obtained when spending propensities are uniformly distributed, which is empirically observed across users of Sarafu. 
Furthermore, we predict that heavy-tailed distributions of the transaction sizes, empirically observed in previous works~\cite{starnini_smurf-based_2021}, emerge when overdispersion 
in the spending propensity of agents is introduced in the model. 
Finally, we show that the conservative nature of circulation provides a signature one-to-one correlation between total inflows and outflows, as seen also in real economic systems~\cite{bacilieri_firm-level_2023}.

\section{Model definition}

We model financial transactions as random walks on temporal networks~\cite{starnini_random_2012, hoffmann_generalized_2012, barrat_modeling_2013}. 
We consider a discrete amount of money $M$, much larger than the network size $N$, $M \gg N$. 
Each unit of money is represented as a single random walker (RW). For instance, one million U.S. Dollars are represented by $M = 10^6$ simultaneous RWs. 
In the network, nodes represent bank accounts, and a directed edge from node $i$ to node $j$ at time $t$ represents a financial transaction between the corresponding bank accounts, whose amount $w_{ij}(t)$ is equal to the number of RWs jumping from node $i$ to node $j$. 
Likewise, the number of RWs present at node $i$ at time $t$ represents the current balance of that node, $m_i(t)$. 
Crucially, RWs are passive: They 
may step 
from one node to some other only when edges are available~\cite{masuda_random_2017}, i.e., money is transferred only through an active temporal edge between two agents. 
Therefore, we model financial transactions as the interplay between network dynamics, representing the directed, instantaneous interactions among the agents of the system, and passive random walk processes, modeling the transfer of funds (i.e., payments) between interacting agents.

We model interactions among accounts as a directed, activity-driven, temporal network with attractiveness~\cite{alessandretti_random_2017, pozzana_epidemic_2017, moinet_generalized_2018}. 
Each node activates by following a renewal process with their activation probability, and fires one directed link to a neighbor, chosen according to their fitness or attractiveness~\cite{caldarelli_scale-free_2002, medo_temporal_2011}. 
Edges are instantaneous, i.e., they have a negligible duration. 
Each node $i$ is characterized by two features: i) time-dependent activity $a_i(t)$, and ii) attractiveness $b_i$. 
The activity $a_i(t)$ represents the probability per unit time that agent $i$ activates at time $t$ since its last activation event. 
This defines the inter-event time distribution $\psi_i(\tau)$, representing the probability of waiting a time $\tau$ between consecutive activations~\cite{karsai_small_2011, moinet_burstiness_2015}. 
While time-independent activity $a_i(t) = a_i$ yields a Poissonian process with exponential inter-event times, non-Poissonian processes produce fat-tailed distributions characteristic of bursty human behavior~\cite{karsai_bursty_2018}. 
We model such burstiness using a Weibull distribution, where the shape parameter $k$ controls the tail: $k = 1$ recovers a Poisson process, while $k < 1$ produces heavy-tailed, bursty dynamics. 
Upon activation, nodes engage in one directed interaction with a selected peer. 
The probability of selecting a certain node $j$ is proportional to their attractiveness $b_j$, assumed to be constant in time.

When an edge is created from node $i$ to node $j$ at time $t$, each RW present at node $i$ jumps independently to node $j$ with a certain probability. 
The probability depends on the spending {propensity }
of node $i$, $s_i \in \left[ s_\mathrm{min}, s_\mathrm{max} \right]$, {representing} 
the {tendency }
of agent $i$ to spend a fraction $s_i$ of their current balance. Upon activation, 
nodes with small spending 
propensities, $s_i \ll 1$, tend to retain most RWs, 
while nodes with large spending 
propensity, $s_i \simeq s_\mathrm{max}$, will send almost all RWs to the target node. 
Heterogeneity in spending propensity is a 
topic commonly explored in asset-exchange models~\cite{chakraborti_statistical_2000, chatterjee_kinetic_2007}, while heterogeneity in money velocity, a related empirical quantity, 
has been noted in bitcoin~\cite{campajola_microvelocity_2022}, Ethereum~\cite{collibus_microvelocity_2025}, and Sarafu~\cite{mattsson_inverse_2023}.

In the simplest version of the payment dynamics, all RWs on node $i$ at time $t$, $m_i(t)$, jump with the same probability $s_i$. 
If a RW does not jump (with probability $1 - s_i$), it remains at node $i$ and waits for the next activation event of node $i$. 
The number of RWs jumping from node $i$ to node $j$ at time $t$, representing the transaction size, $w_{ij}(t)$, will thus be distributed according to a Binomial distribution, $w_{ij}(t) \sim \mathrm{Bin}(m_i(t), s_i)$. 
We consider also a more realistic version of the model, where agents exhibit {greater variance} 
in their spending behavior: The same agent may decide to spend a larger or smaller fraction of their balance. 
We assume that $w_{ij}(t)$ follows a Beta-Binomial distribution, that is $w_{ij}(t) \sim \mathrm{BetaBin}(m_i(t), \xi s_i, \xi (1 - s_i))$, 
where $\xi \in [0, \infty)$ accounts for the precision around the mean value $s_i$. 
For $\xi \to 0$ one has the maximum variance, 
while for $\xi \to \infty$ one recovers the Binomial distribution: $\lim_{\xi \to \infty} \mathrm{BetaBin}(m_i(t), \xi s_i, \xi (1 - s_i)) = \mathrm{Bin}(m_i(t), s_i)$.

Therefore, in the simplest Poissonian formulation (i.e., activity constant in time), the temporal network is characterized by the joint probability distribution $P(a, b)$ of finding a node with activity $a$ and attractiveness $b$ in the population. 
The marginal distributions $f(a)$ and $g(b)$ represent the activity and attractiveness density distributions, respectively. 
If no correlations are present, then $P(a, b) = f(a) \, g(b)$ and activity and attractiveness of nodes are sampled independently. 
Correlations between these two quantities are introduced via a standard bivariate copula on uniform marginals (see Supplementary Material (SM) for details), as done in quantitative finance~\cite{mcneil_quantitative_2015}. 
On the other hand, the payment dynamics is fully characterized by the distribution of the spending propensities in the population, $h(s)$, which we assume independent from the joint activity-fitness distribution, $P(a, b)$, and the precision parameter, $\xi$.

\section{Analytical solution}

We are interested in deriving two key observables in the stationary state: 
The distribution of RWs over the network, $P(m)$, and the distribution of the number of RWs jumping through the edges, $P(w)$. 
The former represents the account balance distribution, 
while the latter represents the transaction size distribution.

We start by calculating the mean number of RWs expected at a node with features $(a,b,s)$, 
which reads (see SM)
\begin{equation}
    \lambda_{a,b,s} = M\pi_{a,b,s} = K_0 \frac{b}{as},
    \label{eq:lambda}
\end{equation}
where $\pi_{a,b,s}$ is the probability of finding a single RW on a node with features $(a, b, s)$ in the steady state, $K_0 \equiv \frac{M}{N \av{b/a}_P \, \av{s^{-1}}_h}$, and $\av{\ldots}_P$, $\av{\ldots}_h$ denote the mean value computed from the distribution $P(a, b)$ and $h(s)$, respectively. 
Note that $\lambda_{a,b,s}$ is proportional to the attractiveness $b$ (determining the number of incoming RWs), and inversely proportional to the activity $a$ and spending propensity $s$  (determining the number of outgoing RWs) of the node. 
Figure~1 of the SM 
shows that Eq.~\eqref{eq:lambda} is validated by numerical simulations of the model and that the derivation of $\pi_{a, b, s}$ is valid for both high ($\xi \to \infty$) and low ($\xi = 1$) spending precision.

The probability of finding $m$ RWs on a randomly chosen node in the network, $P(m)$, in the stationary state, reads
\begin{equation}
    P(m) = \int
    p(m \mid a, b, s) \, F(a, b, s) \, da \, db \, ds,
    \label{eq:general_Pm}
\end{equation}
where $p(m \mid a, b, s)$ is the probability of finding $m$ RWs on a node with features $(a, b, s)$ and $F(a, b, s) = P(a,b) \, h(s)$ is the probability of observing a node with these features. 
On a static network, RWs would be independent, thus $p(m \mid a, b, s)$ would follow a Binomial distribution, $p(m \mid a, b, s) = \mathrm{Bin} \left( M, \pi_{a, b, s} \right)$. 
For large $M$ and small $\pi_{a,b,s}$, the conditional probability becomes a Poisson distribution with mean $\lambda_{a, b, s}$. 
However, on a temporal network, RWs do not move independently. 
While not interacting directly, they accumulate on a given node until it activates, at which point each RW may jump with probability $s$. 
When $s$ is large, RWs tend to 
jump together. 
In the limit of small $s$, instead, just a few RWs jump, and we recover the independence assumption. We will treat first the simpler small $s$ case and next the general $s$ case.

In the following, we assume time-independent activity $a_i=a_i(t)$, perfect correlation between activity and attractiveness, $b = a$, and we focus on a general power-law  distribution of spending propensities of the form $h(s) \sim s^\sigma$, where $\sigma=0$ represents a uniform distribution.
In this case, $P(m)$ does not depend on the joint probability $P(a, b)$ anymore, as they factor out from the integral in Eq.~\eqref{eq:general_Pm}, and $\lambda_s = K_0/s$, where now $K_0 = \frac{M}{N \av{s^{-1}}_h}$.
Note that the minimum $s_\mathrm{min}$ cannot go to zero to avoid a divergence in $\lambda_s$. 
Indeed, if a node has a zero spending propensity, all RWs would never jump and would accumulate on that node. 
Without loss of generality, we define $s_\mathrm{min} = s_\mathrm{max}/N$.

\begin{figure}[tbp]
    \centering
    \includegraphics[width=0.9\columnwidth]{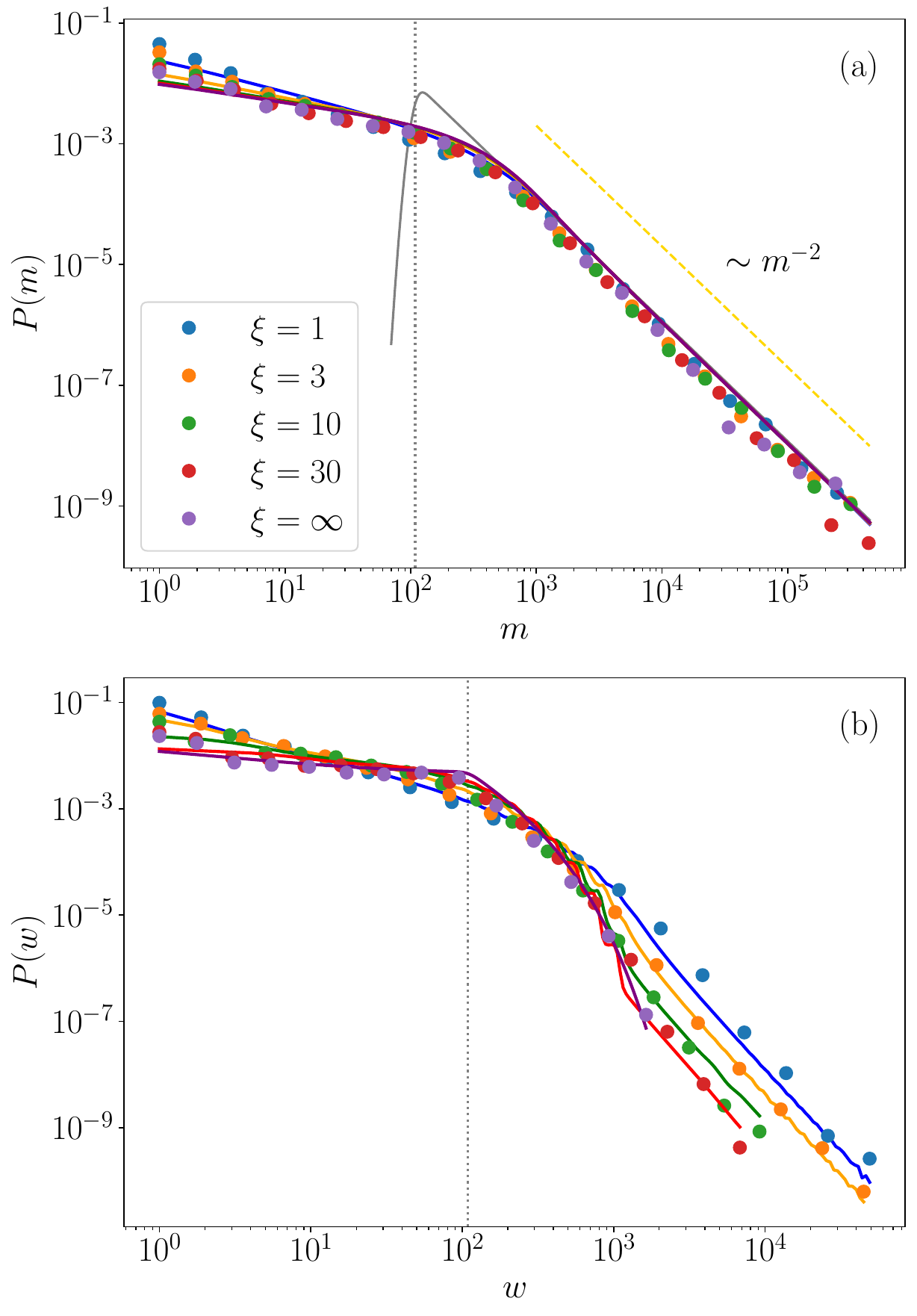}
    \caption{
    Probability distributions $P(m)$ (a) and $P(w)$ (b) for different values of the precision parameter $\xi$. 
    Yellow dashed line in panel (a) shows a power-law scaling $P(m) \sim m^{-2}$. 
    Vertical dotted lines correspond to $m = K_0/s_\mathrm{max}$ (a) and $w = K_0$ (b). 
    Gray line in panel (a) corresponds to the analytical solution of $P(m)$ obtained in the limit of $\xi \to \infty$ and small $s$, see Eq.~(18) of the SM. 
    Solid lines correspond to numerical integrations of Eq.~(29) (a) and Eq.~(43) (b) of the SM. 
    We used $N = 10^4$, $M = 10^7$, $\sigma = 0$, $s_\mathrm{max} = 1.0$, and perfect correlation $b = a$.
    }
    \label{fig:Pm_Pw}
\end{figure}

For $m \gg K_0/s_\mathrm{max}$, and if $\sigma \in \mathbb{Z}$, Eq.~\eqref{eq:general_Pm} leads to the asymptotic behavior (see SM)
\begin{equation}
    P(m) \sim m^{-(\sigma + 2)}.
    \label{eq:decay_Pm}
\end{equation}
In particular, for $\sigma = 0$---i.e., uniformly distributed spending propensities---we obtain a power-law decay $P(m) \sim m^{-2}$, which has also been found in the asset-exchange model~\cite{chatterjee_kinetic_2007}. 
Figure~\ref{fig:Pm_Pw}(a) validates this asymptotic behavior by numerical simulations with $\sigma = 0$, 
for different values of $\xi$. 
For different values of $\sigma$, see Fig.~4 of the SM. 
Note that, for $m > K_0/s_\mathrm{max}$, the $P(m)$ distribution is well reproduced by the analytical prediction given by Eq.~(18) of the SM, shown as a gray line in Fig.~\ref{fig:Pm_Pw}(a). 
Furthermore, Eq.~\eqref{eq:decay_Pm} is valid for any distribution $h(s) \sim s^{\sigma}$ as long as the minimum $s_\mathrm{min} = s_\mathrm{max}/N$ vanishes in the thermodynamic limit $N \to \infty$, as the tail of the $P(m)$ is determined by nodes with small $s$, for which we can approximate $p(m \mid s)$ to a Poisson distribution. 
This means that the power-law scaling of the balance distribution, $P(m) \sim m^{-2}$, 
can arise even if the spending rate across the population is uniform with $s_\mathrm{max} = 1$, due to a few agents spending a small fraction of their balance per activation and thus accumulating funds.

However, for non-vanishing values of $s$, RWs cannot be assumed independent; the conditional probability of finding $m$ RWs on a node with spending rate $s$, $p(m \mid s)$, is no longer a Poisson distribution. 
When many RWs jump from a node, they all end up at the same destination, introducing correlation in their co-location. 
Modeling this correlation, 
the Poisson distribution 
becomes a negative binomial distribution, $p(m \mid s) = \mathrm{NB} \left( r, \frac{r}{r + \lambda_s} \right)$, 
with the same mean $\lambda_s$ (see SM for details). 
The parameter $r$ accounts for the correlations induced by co-jumping, which is set to $r = \frac{1 - \av{s}_h}{s}$, in line with previous findings on wealth-exchange models~\cite[see A2-model]{patriarca_basic_2010}. 
For small $s$, the parameter $r$ diverges and the negative binomial distribution becomes a Poisson distribution, thus recovering the uncorrelated case. 
Figure~2 of the SM shows that numerical simulations are in very good agreement with this parametrization of the $p(m \mid s)$. 
Under this assumption, one can obtain an expression for the $P(m)$, which can be integrated numerically, see SM. 
In the Beta-Binomial model, one has a further source of dispersion in the $p(m \mid s)$ distribution due to the precision parameter $\xi$. 
In this case, we define an effective parameter $r_{\mathrm{eff}} = (r^{-1} + \xi^{-1})^{-1}$ accounting for the overall correlations, see SM for details.
Figure~\ref{fig:Pm_Pw}(a) shows that numerical simulations nicely agree with the numerical integration of Eq.~(29) of the SM, for $\sigma = 0$ (uniform spending rate distribution) and for different values of $\xi$. 
Therefore, we can analytically predict the behavior of $P(m)$ for large $m$ and obtain it numerically for small values of $m$, for a generic form of the spending propensity distribution $h(s)$.

The probability of observing $w$ RWs jumping during a randomly chosen activation event, $P(w)$, reads
\begin{equation}
    P(w) = \sum_{m = w}^{M} \int_{s_\mathrm{\min}}^{s_\mathrm{\max}} p(w \mid m, s) \, p(m \mid s) \, h(s) \, ds,
    \label{eq:reduced_Pw}
\end{equation}
where $p(w \mid m, s)$ is the probability of observing $w$ RWs jumping from a node with $m$ RWs and spending propensity $s$. 
The conditional probability $p(w \mid m, s)$ is what marks the difference between high ($\xi \to \infty$) and low (finite $\xi$) spending precision. 
In the former case, we have $p(w \mid m, s) = \mathrm{Bin}(m, s)$. 
Under the independence assumption, i.e., small $s$, 
$p(m \mid s) = \mathrm{Pois} \left( \lambda_s \right)$. Therefore, the distribution takes the form (see SM)
\begin{equation}
    P(w) = \frac{\left( K_0 \right)^w \, e^{-K_0}}{w!}.
    \label{eq:Pw_Poisson}
\end{equation}
Figure~5 of the SM shows that Eq.~\eqref{eq:Pw_Poisson} provides a good approximation of $P(w)$ for low $s_\mathrm{max}$. 
However, for general $s$, co-location correlations arise and the probability $p(m \mid s)$ is better approximated by a negative binomial. 
In this case, we can obtain an expression for the $P(w)$ to be integrated numerically, see SM. 

In the case of finite $\xi$, instead, we have $p(w \mid m, s, \xi) = \mathrm{BetaBin}(m, \xi s, \xi (1 - s))$. 
For general $s$, one can obtain an expression for the $P(w)$ by compounding the binomial-beta mixture and the Poisson-gamma mixture representing the beta-binomial and negative binomial distributions, respectively, see SM. 
Figure~\ref{fig:Pm_Pw}(b) shows that numerical simulations of the model are in good agreement with the numerical integration of Eq.~(43) of the SM, for $\sigma = 0$ and for different values of $\xi$. 
Crucially, while 
for $\xi \to \infty$ we obtain 
a Poissonian decay in the transaction size distribution $P(w)$ (as predicted by Eq.~\eqref{eq:Pw_Poisson}), 
for finite $\xi$ the model produces a heavy-tailed distribution. 
This behavior is especially pronounced for low values of $\xi$, corresponding to high dispersion around the mean spending propensity.

\section{Empirical data}

We calibrate our model on a dataset of financial transactions in Sarafu, a stand-alone digital community currency in Kenya, from January 2020 to June 2021. 
Grassroots Economics Foundation, the Kenyan nonprofit organization that operates Sarafu, has published pseudonymized system records via the UK Data Service~\cite{ruddick_sarafu_2021}. The transaction dataset is comprehensive and thoroughly documented~\cite{mattsson_sarafu_2022}. 
One Sarafu corresponded to approximately 2.4 U.S. cents in 2021 terms (adjusted for purchasing power parity). 
We filter out low-active users (with either zero incoming or outgoing transactions) and small transactions (less than one Sarafu), see SM.
We reconstruct the balance history for each user, starting from the final balance and going backward in time, see SM. This process led to inconsistencies for 4 users, 
which are discarded. 
The resulting dataset comprises approximately 400,000 transactions among around 24,000 users, totaling approximately 300 million Sarafu, equivalent to around 2.8 million USD.

\section{Model calibration}

We calibrate the network dynamics by assigning activity and attractiveness to the agents.
The number of outgoing ($n^\mathrm{out}$) and incoming ($n^\mathrm{in}$) transactions per agent are power-law distributed (see Fig.~8 of the SM), remarkably stable over time (see Figs.~9, 10, and Table~1 of the SM), and strongly correlated (see Fig.~11 of the SM). 
We model this correlation by fitting the joint $P(n_i^\mathrm{in}, n_i^\mathrm{out})$ distribution to a Joe copula, obtaining a very good approximation (see Fig.~13 and Table~2 of the SM). 
Next, we calibrate the activity rates to have an expected number of outgoing transactions per unit of time in the numerical simulations equal to the one observed in the Sarafu dataset. 
Within the Poissonian formulation, this simply corresponds to an activity rate $a_i$ equal to the number of outgoing transactions $n_i^\mathrm{out}$ of agent $i$ divided by the length of the time window $T$ of the data. 
Within the non-Poissonian formulation, we first select a global shape parameter $k$ of the Weibull distribution for all agents, corresponding to a bursty behavior. 
This assumption is justified by the high degree of similarity in individual inter-event time distributions $P(\tau)$ observed across the dataset, see Fig.~12 of the SM. 
Then, we calibrate the scale parameter of the Weibull distribution for each agent to obtain the expected number of outgoing transactions sampled in the model, see Eq.~(46) of the SM. 
The attractiveness of the agents is calibrated by simply dividing the number of incoming transactions $n_i^\mathrm{in}$ of agent $i$ by the total number of incoming transactions in the system, $b_i = n_i^\mathrm{in} / \sum_\ell n_\ell^\mathrm{in}$.

\begin{figure}[tbp]
    \centering
    \begin{subfigure}{0.45\textwidth}
        \centering
        \includegraphics[width=\textwidth]{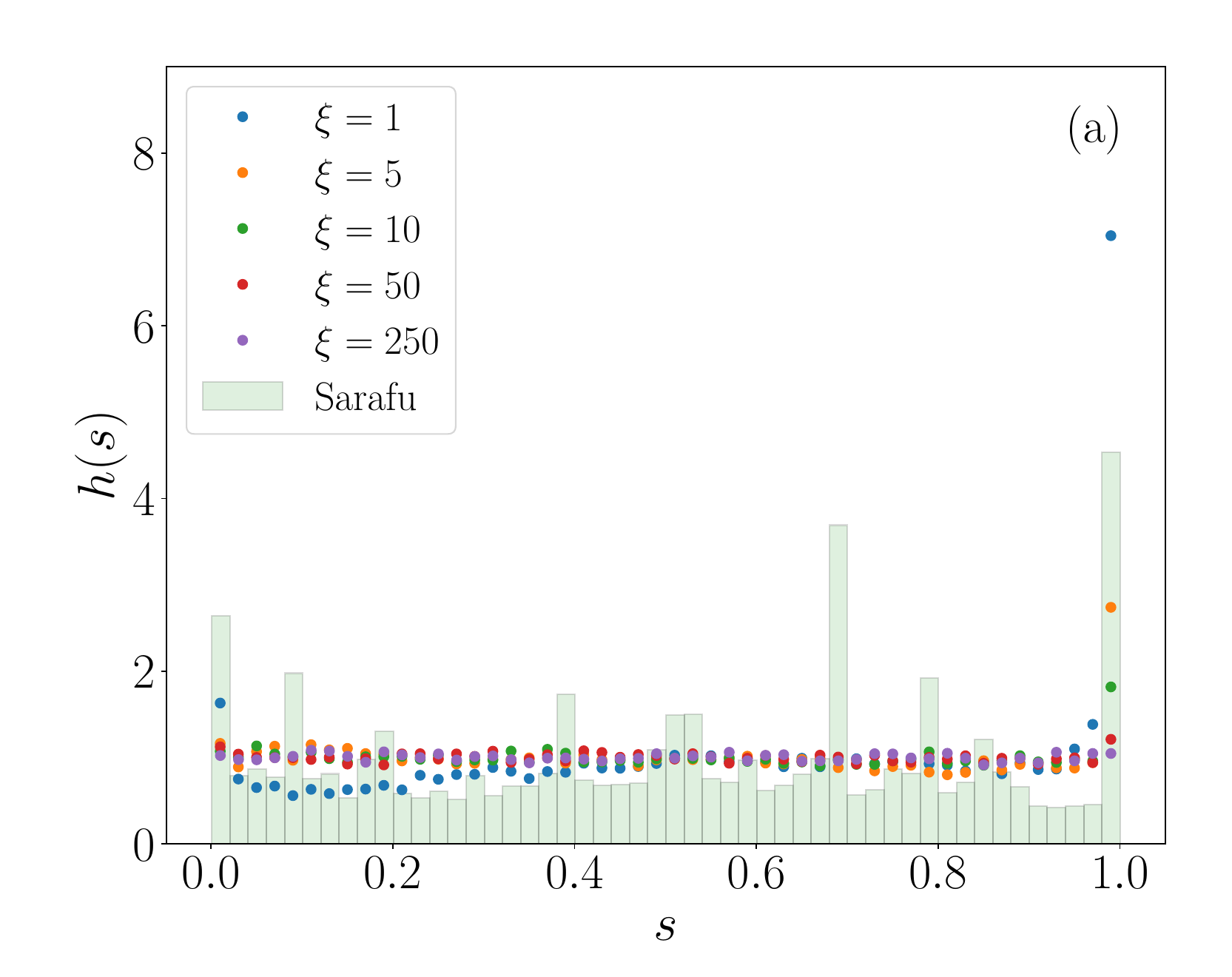}
        \label{fig:}
    \end{subfigure} %
\\[-18pt]      
    \begin{subfigure}{0.45\textwidth}
        \centering
        \includegraphics[width=\textwidth]{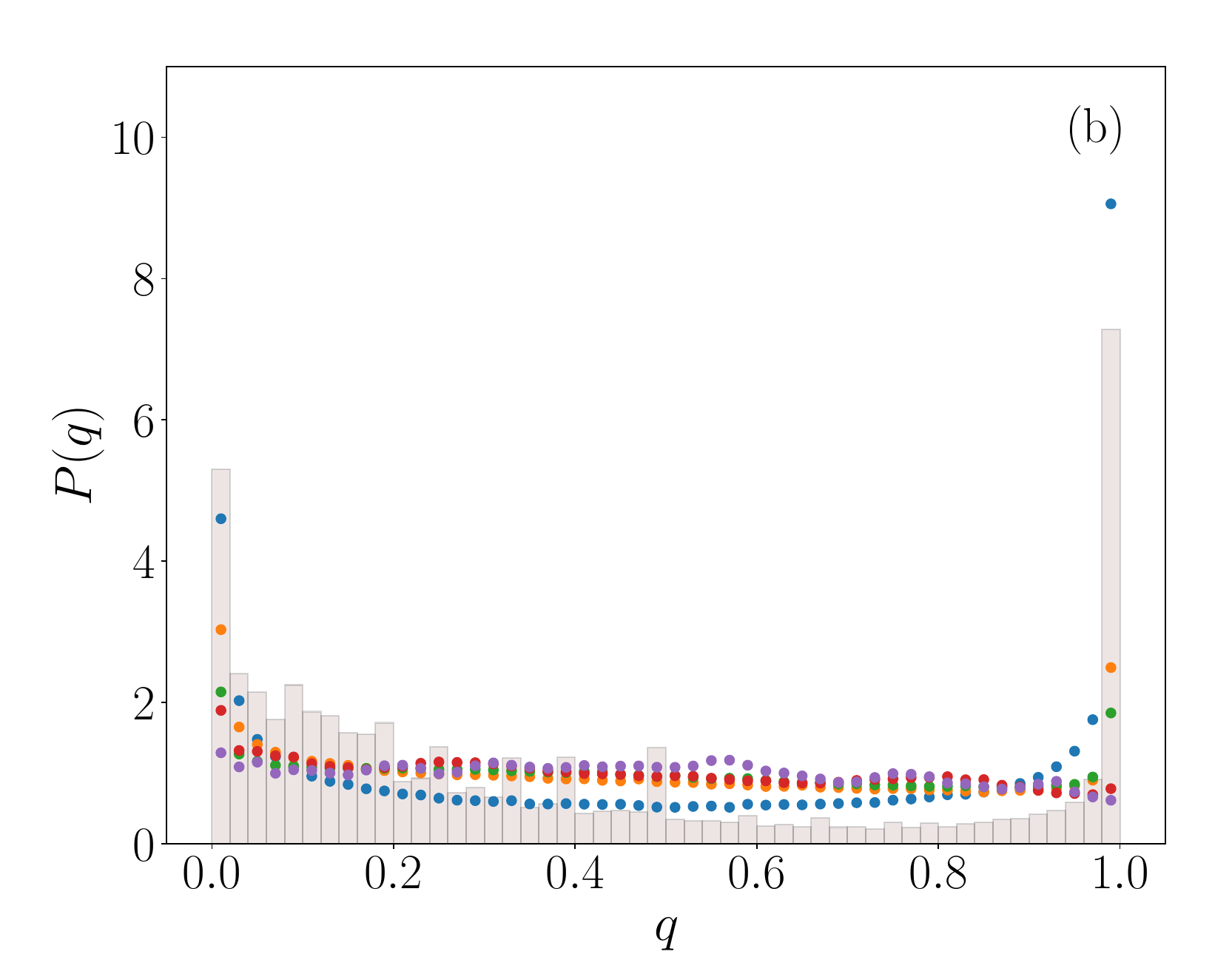}
        \label{fig:}
    \end{subfigure}
    \caption{Probability distribution of time-average spending propensities $h(s)$ (a) and spending fractions $P(q)$ (b), observed in the data (bars) and obtained from simulations of the model (points), for different values of the precision parameter $\xi$ and a shape parameter $k = 0.75$.}
    \label{fig:spending_fractions_and_parameters}
\end{figure}

Finally, we calibrate the payment dynamics. 
In the model, each agent $i$ is characterized by their spending propensity $s_i$ and the precision around this average value, parametrized by $\xi$. 
These two quantities can be measured in the data by the fraction of money spent in a single transaction, $q_i(t) = w_{ij}(t) / m_i(t)$. 
The spending propensity of a user $i$, $s_i$, is computed by averaging the spending fractions over the whole time series, that is $s_i = \av{q_i(t)}_T$. 
Figure~\ref{fig:spending_fractions_and_parameters}(a) shows that the empirical spending propensities of agents are remarkably near to uniformly distributed across the population. 
Therefore, 
we choose $h(s) \sim \mathcal{U}_{\left[ s_\mathrm{min}, s_\mathrm{max} \right]}$, corresponding to $\sigma = 0$ in the model. 
However, Fig.~\ref{fig:spending_fractions_and_parameters}(b) shows that the distribution of spending fractions over the population and over time, $P(q)$, has a typical U-shape, with peaks at the extremes $q = 0$ and $q = 1$. 
This implies that, while the average spending propensity of a user is uniformly distributed across the population,  jump probabilities also involve variance around this mean at the level of the individual. 
One can see that this behavior is well reproduced by the model for low values of the precision parameter $\xi$, in particular for $\xi = 1$ and $\xi = 5$. 
Therefore, 
we set $\sigma = 0$ for $h(s)$ (uniform distribution) and $\xi = 1$ (low spending precision). 

\section{Numerical results}

We run numerical simulations of the model with $N = 25000$ agents. 
In order to ensure that the stationary state has been reached, we run 6 million transactions and analyze the last $5 \times 10^5$.
Both the population size and the total number of transactions are of the same order of magnitude as those in the Sarafu dataset. 
The distribution of initial balances (number of RWs) is constant, $m_i(t = 0) = m_0 = 100$, for a total number of RWs equal to $M =  m_0 \, N$.

\begin{figure}[tbp]
    \centering
    \includegraphics[width=0.9\columnwidth]{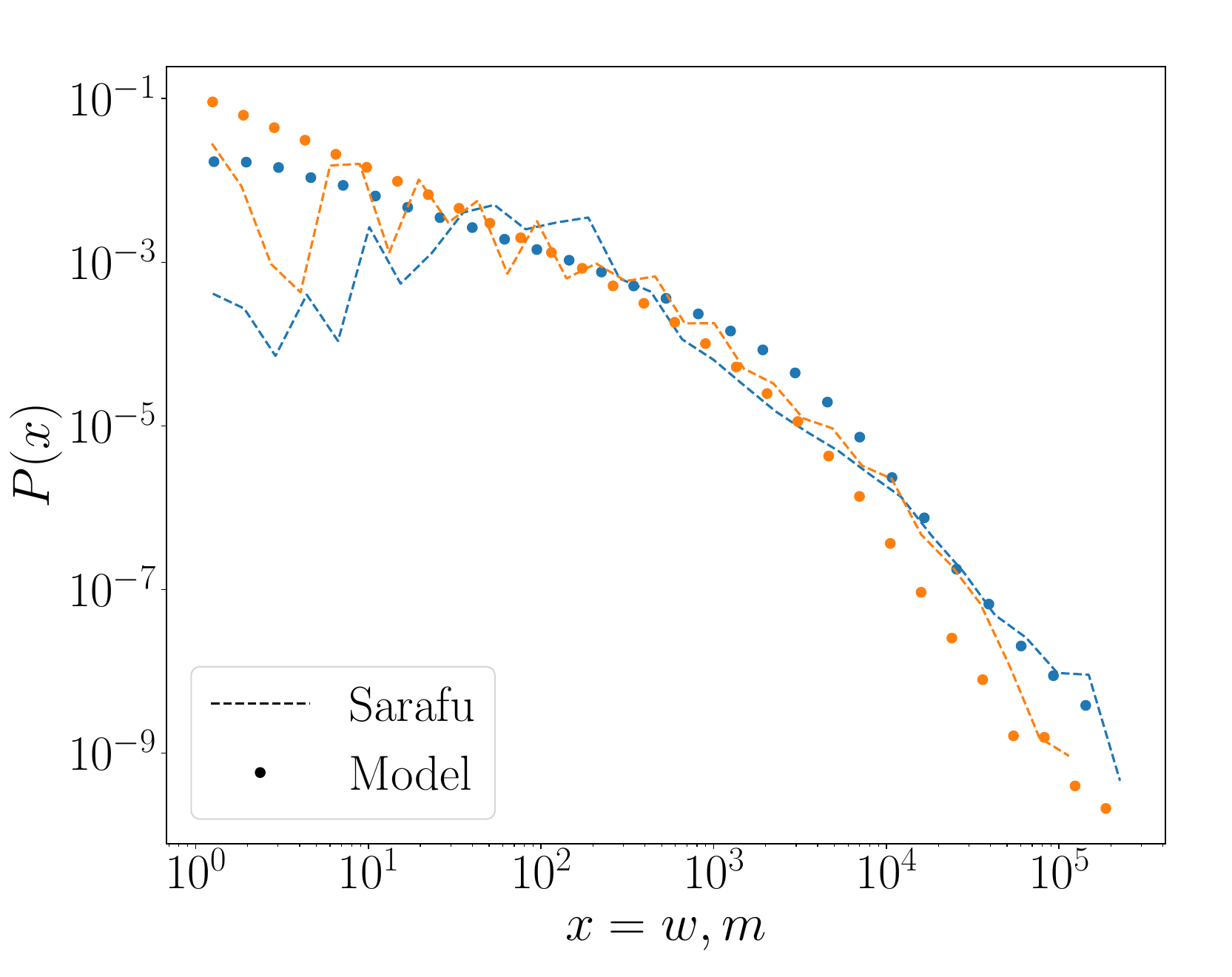}
    \caption{
    Probability distributions of balances $P(m)$ (blue) and transaction sizes $P(w)$ (orange), observed in the data (dashed lines) and obtained from numerical simulations of the model with $\xi = 1$ and $k = 0.75$ (points).
    }
    \label{fig:balance_trans_distr}
\end{figure}

We measure the probability distribution of RWs over the network in the stationary state, $P(m)$, corresponding to the balance distribution across Sarafu accounts. 
Figure~\ref{fig:balance_trans_distr} shows the balance distribution $P(m)$ observed in the data and obtained from simulations of the model with $k = 0.75$, i.e., non-Poissonian activation pattern, and $\xi = 1$, i.e., low spending precision. 
One can see that the numerical simulations are in very good agreement with the empirical data. 
Indeed, $P(m)$ shows a fat tail compatible with a power-law decay 
for large $m$, as predicted by Eq.~\eqref{eq:decay_Pm}. 
Figure~\ref{fig:balance_trans_distr} also shows the transaction size distribution $P(w)$ observed in the data and obtained from simulations. 
Again, one can observe a good agreement between the empirical $P(w)$ and that obtained from numerical simulations.

The empirical balance distribution $P(m)$ is well reproduced by the model also for $k = 1$ (Poissonian activation pattern), and both small and large $\xi$ (low and high spending precision, respectively), see Fig.~15 of the SM. 
Instead, the empirical transaction size distribution $P(w)$ is well reproduced by the model for $k = 1$, but only for small $\xi$. 
Indeed, a Poissonian decay in $P(w)$ is expected for $\xi \to \infty$, see Eq.~\eqref{eq:Pw_Poisson}. 
Therefore, we conclude that bursty activation patterns do not have a significant impact on the model behavior, while individual-level variance 
in spending behavior---modeled by $\xi$---is directly responsible for the heavy-tailed nature of the transaction size distribution, also observed within large-scale payment systems~\cite[][see Fig. 2(b)]{starnini_smurf-based_2021}. 
Note that the model also captures the distribution of the average balance $\av{m}$ and average transaction size $\av{w}$ per Sarafu user, 
for any values of $\xi$ and $k$, see Fig.~16 of the SM.

\begin{figure}
    \centering
    \includegraphics[width=0.95\linewidth]{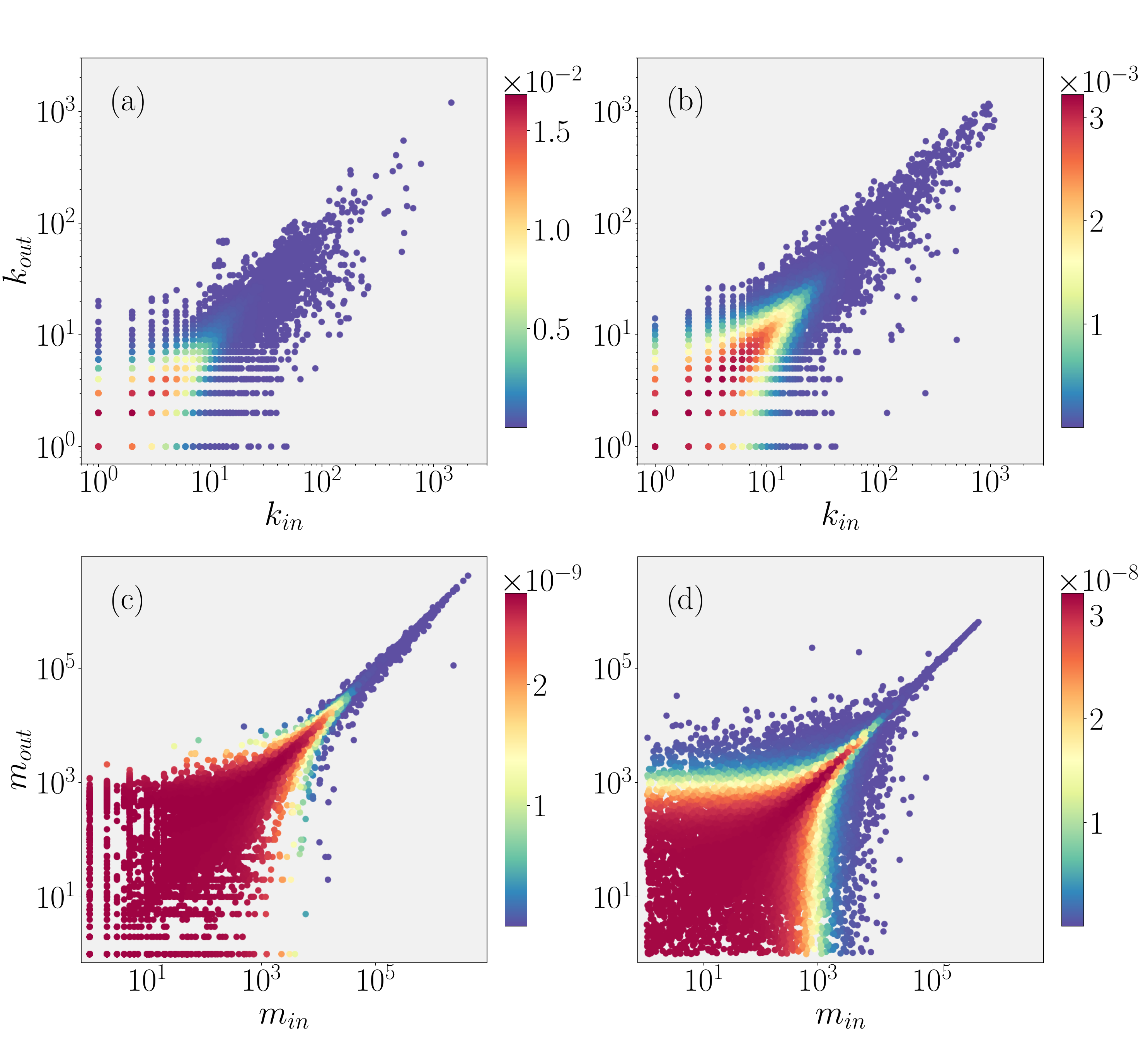}
    \caption{
    Joint probability distribution of in-degree ($k_\mathrm{in})$ and out-degree ($k_\mathrm{out})$ of individuals, observed in the dataset (a) and the model with $\xi = 1$ and $k = 0.75$ (b). 
    Joint probability distribution of total money received ($m_\mathrm{in})$ and total money spent ($m_\mathrm{out}$) by individuals, observed in the dataset (c) and the model with $\xi = 1$ and $k = 0.75$ (d). 
    Each point in the plots represents a single user. 
    The color represents the density value of users estimated through a kernel density estimation using Gaussian kernels.
    }
    \label{fig:k_inout_m_inout}
\end{figure}

The model is able to properly reproduce the topological properties of the underlying transaction network, as well as correlations in 
transaction volumes. 
The joint activity-attractiveness distribution $P(a, b)$ of the empirical data is recovered by the model, see Fig.~14 of the SM. 
The in-degree and out-degree distributions of the time-aggregated transaction network obtained from the model, $P \left( k_\mathrm{in} \right)$ and $P \left( k_\mathrm{out} \right)$, respectively, are in good agreement with the empirical distributions, showing a behavior compatible with a power-law decay, see Fig.~17 of the SM. 
Furthermore, the model can also reproduce the correlation between in- and out-degrees of the nodes, shown by the joint probability distribution $P \left( k_\mathrm{in}, k_\mathrm{out} \right)$ in Fig.~\ref{fig:k_inout_m_inout}(a) and (b). 
The same applies for the correlation between the total in- and out-strength of the nodes, corresponding to the total amount of money sent and received, respectively, shown by the joint probability distribution $P \left( m_\mathrm{in}, m_\mathrm{out} \right)$ in Fig.~\ref{fig:k_inout_m_inout}(c) and (d).

\section{Conclusions}

We have shown that financial transactions can be effectively modeled as a system of simultaneous random walks on activity-driven temporal networks. 
This framework reproduces the heavy-tailed distributions observed in empirical data for both account balances $P(m)$ and transaction sizes $P(w)$. 
In our model,
while balance heterogeneity stems from heterogeneity in payment dynamics, a feature of several empirical payment systems~\cite{campajola_microvelocity_2022, collibus_microvelocity_2025, mattsson_inverse_2023}, the recently observed heavy-tailed nature of transaction sizes~\cite{starnini_smurf-based_2021} 
emerges from 
individual-level variance 
in spending behavior. 
By respecting the fundamental constraint of fund conservation, the model naturally recovers correlations between inflow and outflow volumes empirically observed in the Sarafu dataset and the literature~\cite{bacilieri_firm-level_2023}. 
Future research should introduce memory into the activity-driven framework to account for agents repeatedly sending money to the same individuals. 
Furthermore, one could model open payment systems by explicitly accounting for deposits, withdrawals, and token creation at system boundaries.

\section{Data availability}
The Sarafu 2020-21 records~\cite{ruddick_sarafu_2021, mattsson_sarafu_2022} are available via the UK Data Service (UKDS) under their End User License, which stipulates suitable data-privacy protections. The data is available for download from the UKDS ReShare repository (\url{reshare.ukdataservice.ac.uk/855142/}) to users registered with the UKDS (\url{beta.ukdataservice.ac.uk/myaccount/credentials}).

\begin{acknowledgments}
    J.~O. acknowledges funding from the Program for Units of Excellence in R\&D Mar\'ia de Maeztu (CEX2021-001164-M,  MICIU/AEI/10.13039/501100011033). 
    M.~S. acknowledges support from Grants No. RYC2022-037932-I and CNS2023-144156 funded by MCIN/AEI/10.13039/501100011033 and the European Union NextGenerationEU/PRTR. 
    We thank Marco Maggiora, Enrico Scalas, Bruno Coutinho, Ann Wills, Rangel Baldasso, and Kleber Oliveira for useful discussion.
\end{acknowledgments}

\bibliography{references}

@article{alessandretti_random_2017,
	title = {Random walks on activity-driven networks with attractiveness},
	volume = {95},
	url = {https://link.aps.org/doi/10.1103/PhysRevE.95.052318},
	doi = {10.1103/PhysRevE.95.052318},
	number = {5},
	urldate = {2025-12-10},
	journal = {Phys. Rev. E},
	author = {Alessandretti, Laura and Sun, Kaiyuan and Baronchelli, Andrea and Perra, Nicola},
	month = may,
	year = {2017},
	pages = {052318},
}

@article{pozzana_epidemic_2017,
	title = {Epidemic spreading on activity-driven networks with attractiveness},
	volume = {96},
	url = {https://link.aps.org/doi/10.1103/PhysRevE.96.042310},
	doi = {10.1103/PhysRevE.96.042310},
	number = {4},
	urldate = {2025-12-10},
	journal = {Phys. Rev. E},
	author = {Pozzana, Iacopo and Sun, Kaiyuan and Perra, Nicola},
	month = oct,
	year = {2017},
	pages = {042310},
}

@article{moinet_generalized_2018,
	title = {Generalized voterlike model on activity-driven networks with attractiveness},
	volume = {98},
	url = {https://link.aps.org/doi/10.1103/PhysRevE.98.022303},
	doi = {10.1103/PhysRevE.98.022303},
	number = {2},
	urldate = {2025-12-10},
	journal = {Phys. Rev. E},
	author = {Moinet, Antoine and Barrat, Alain and Pastor-Satorras, Romualdo},
	month = aug,
	year = {2018},
	pages = {022303},
}

@article{mattsson_trajectories_2021,
	title = {Trajectories through temporal networks},
	volume = {6},
	issn = {2364-8228},
	url = {https://doi.org/10.1007/s41109-021-00374-7},
	doi = {10.1007/s41109-021-00374-7},
	number = {1},
	urldate = {2025-12-10},
	journal = {Appl Netw Sci},
	author = {Mattsson, Carolina E. S. and Takes, Frank W.},
	month = may,
	year = {2021},
	pages = {35},
}

@article{perra_activity_2012,
	title = {Activity driven modeling of time varying networks},
	volume = {2},
	copyright = {2012 The Author(s)},
	issn = {2045-2322},
	url = {http://www.nature.com/articles/srep00469},
	doi = {10.1038/srep00469},
	number = {1},
	urldate = {2021-09-13},
	journal = {Sci Rep},
	author = {Perra, N. and Gonçalves, B. and Pastor-Satorras, R. and Vespignani, A.},
	month = jun,
	year = {2012},
	pages = {469},
}

@article{starnini_random_2012,
	title = {Random walks on temporal networks},
	volume = {85},
	url = {https://link.aps.org/doi/10.1103/PhysRevE.85.056115},
	doi = {10.1103/PhysRevE.85.056115},
	number = {5},
	urldate = {2025-12-11},
	journal = {Phys. Rev. E},
	author = {Starnini, Michele and Baronchelli, Andrea and Barrat, Alain and Pastor-Satorras, Romualdo},
	month = may,
	year = {2012},
	pages = {056115},
}

@article{starnini_topological_2013,
	title = {Topological properties of a time-integrated activity-driven network},
	volume = {87},
	url = {https://link.aps.org/doi/10.1103/PhysRevE.87.062807},
	doi = {10.1103/PhysRevE.87.062807},
	number = {6},
	urldate = {2021-10-01},
	journal = {Phys. Rev. E},
	author = {Starnini, Michele and Pastor-Satorras, Romualdo},
	month = jun,
	year = {2013},
	pages = {062807},
}

@misc{mattsson_inverse_2023,
	title = {Inverse estimation of the transfer velocity of money},
	url = {http://arxiv.org/abs/2209.01512},
	doi = {10.48550/arXiv.2209.01512},
	urldate = {2023-08-14},
	publisher = {arXiv},
	author = {Mattsson, Carolina E. S. and Luedtke, Allison and Takes, Frank W.},
	month = jul,
	year = {2023},
	note = {arXiv:2209.01512 [physics, q-fin]},
}

@article{bouchaud_wealth_2000,
	title = {Wealth condensation in a simple model of economy},
	volume = {282},
	issn = {0378-4371},
	url = {https://www.sciencedirect.com/science/article/pii/S0378437100002053},
	doi = {10.1016/S0378-4371(00)00205-3},
	number = {3},
	urldate = {2026-02-02},
	journal = {Physica A: Statistical Mechanics and its Applications},
	author = {Bouchaud, Jean-Philippe and Mézard, Marc},
	month = jul,
	year = {2000},
	pages = {536--545},
}

@article{dragulescu_statistical_2000,
	title = {Statistical mechanics of money},
	volume = {17},
	issn = {1434-6036},
	url = {https://doi.org/10.1007/s100510070114},
	doi = {10.1007/s100510070114},
	number = {4},
	urldate = {2025-12-10},
	journal = {Eur. Phys. J. B},
	author = {Drăgulescu, A. and Yakovenko, V.M.},
	month = oct,
	year = {2000},
	pages = {723--729},
}

@article{partridge_bank_2023,
	chapter = {Business},
	title = {Bank of {Ireland} glitch let customers withdraw money they didn’t have},
	issn = {0261-3077},
	url = {https://www.theguardian.com/business/2023/aug/16/bank-of-ireland-apologises-after-it-glitch-let-customers-withdraw-money-they-didnt-have},
	urldate = {2024-03-22},
	journal = {The Guardian},
	author = {Partridge, Joanna},
	month = aug,
	year = {2023},
}

@article{rivlin_why_2015,
	chapter = {Magazine},
	title = {Why {New} {Orleans}’s {Black} {Residents} {Are} {Still} {Underwater} {After} {Katrina}},
	issn = {0362-4331},
	url = {https://www.nytimes.com/2015/08/23/magazine/why-new-orleans-black-residents-are-still-under-water-after-katrina.html},
	urldate = {2020-09-14},
	journal = {The New York Times},
	author = {Rivlin, Gary},
	month = aug,
	year = {2015},
}

@article{chakraborti_statistical_2000,
	title = {Statistical mechanics of money: how saving propensity affects its distribution},
	volume = {17},
	issn = {1434-6036},
	shorttitle = {Statistical mechanics of money},
	url = {https://doi.org/10.1007/s100510070173},
	doi = {10.1007/s100510070173},
	number = {1},
	urldate = {2026-01-29},
	journal = {Eur. Phys. J. B},
	author = {Chakraborti, A. and Chakrabarti, B.K.},
	month = sep,
	year = {2000},
	pages = {167--170},
}

@article{garlaschelli_structure_2005,
	series = {Market {Dynamics} and {Quantitative} {Economics}},
	title = {Structure and evolution of the world trade network},
	volume = {355},
	issn = {0378-4371},
	url = {https://www.sciencedirect.com/science/article/pii/S0378437105002852},
	doi = {10.1016/j.physa.2005.02.075},
	number = {1},
	urldate = {2026-02-03},
	journal = {Physica A: Statistical Mechanics and its Applications},
	author = {Garlaschelli, Diego and Loffredo, Maria I.},
	month = sep,
	year = {2005},
	pages = {138--144},
}

@article{masuda_random_2017,
	title = {Random walks and diffusion on networks},
	volume = {716-717},
	issn = {0370-1573},
	url = {https://www.sciencedirect.com/science/article/pii/S0370157317302946},
	doi = {10.1016/j.physrep.2017.07.007},
	urldate = {2025-12-10},
	journal = {Physics Reports},
	author = {Masuda, Naoki and Porter, Mason A. and Lambiotte, Renaud},
	month = nov,
	year = {2017},
	pages = {1--58},
}

@article{hoffmann_generalized_2012,
	title = {Generalized master equations for non-{Poisson} dynamics on networks},
	volume = {86},
	url = {https://link.aps.org/doi/10.1103/PhysRevE.86.046102},
	doi = {10.1103/PhysRevE.86.046102},
	number = {4},
	urldate = {2025-12-11},
	journal = {Phys. Rev. E},
	author = {Hoffmann, Till and Porter, Mason A. and Lambiotte, Renaud},
	month = oct,
	year = {2012},
	pages = {046102},
}

@article{barrat_modeling_2013,
	title = {Modeling {Temporal} {Networks} {Using} {Random} {Itineraries}},
	volume = {110},
	url = {https://link.aps.org/doi/10.1103/PhysRevLett.110.158702},
	doi = {10.1103/PhysRevLett.110.158702},
	number = {15},
	urldate = {2025-12-11},
	journal = {Phys. Rev. Lett.},
	author = {Barrat, Alain and Fernandez, Bastien and Lin, Kevin K. and Young, Lai-Sang},
	month = apr,
	year = {2013},
	pages = {158702},
}

@article{caldarelli_scale-free_2002,
	title = {Scale-{Free} {Networks} from {Varying} {Vertex} {Intrinsic} {Fitness}},
	volume = {89},
	url = {https://link.aps.org/doi/10.1103/PhysRevLett.89.258702},
	doi = {10.1103/PhysRevLett.89.258702},
	number = {25},
	urldate = {2025-12-11},
	journal = {Phys. Rev. Lett.},
	author = {Caldarelli, G. and Capocci, A. and De Los Rios, P. and Muñoz, M. A.},
	month = dec,
	year = {2002},
	pages = {258702},
}

@article{medo_temporal_2011,
	title = {Temporal {Effects} in the {Growth} of {Networks}},
	volume = {107},
	url = {https://link.aps.org/doi/10.1103/PhysRevLett.107.238701},
	doi = {10.1103/PhysRevLett.107.238701},
	number = {23},
	urldate = {2025-12-11},
	journal = {Phys. Rev. Lett.},
	author = {Medo, Matúš and Cimini, Giulio and Gualdi, Stanislao},
	month = dec,
	year = {2011},
	pages = {238701},
}

@article{moinet_burstiness_2015,
	title = {Burstiness and {Aging} in {Social} {Temporal} {Networks}},
	volume = {114},
	url = {https://link.aps.org/doi/10.1103/PhysRevLett.114.108701},
	doi = {10.1103/PhysRevLett.114.108701},
	number = {10},
	urldate = {2024-12-16},
	journal = {Phys. Rev. Lett.},
	author = {Moinet, Antoine and Starnini, Michele and Pastor-Satorras, Romualdo},
	month = mar,
	year = {2015},
	pages = {108701},
}

@book{karsai_bursty_2018,
	address = {Cham},
	series = {{SpringerBriefs} in {Complexity}},
	title = {Bursty {Human} {Dynamics}},
	copyright = {http://www.springer.com/tdm},
	isbn = {978-3-319-68538-0 978-3-319-68540-3},
	url = {http://link.springer.com/10.1007/978-3-319-68540-3},
	urldate = {2025-12-12},
	publisher = {Springer International Publishing},
	author = {Karsai, Márton and Jo, Hang-Hyun and Kaski, Kimmo},
	year = {2018},
	doi = {10.1007/978-3-319-68540-3},
}

@article{ispolatov_wealth_1998,
	title = {Wealth distributions in asset exchange models},
	volume = {2},
	copyright = {1998 EDP Sciences, Springer-Verlag},
	issn = {1434-6036},
	url = {https://link.springer.com/article/10.1007/s100510050249},
	doi = {10.1007/s100510050249},
	number = {2},
	urldate = {2026-02-02},
	journal = {Eur. Phys. J. B},
	author = {Ispolatov, S. and Krapivsky, P. L. and Redner, S.},
	month = mar,
	year = {1998},
	pages = {267--276},
}

@article{patriarca_basic_2010,
	title = {Basic kinetic wealth-exchange models: common features and open problems},
	volume = {73},
	issn = {1434-6036},
	shorttitle = {Basic kinetic wealth-exchange models},
	url = {https://doi.org/10.1140/epjb/e2009-00418-6},
	doi = {10.1140/epjb/e2009-00418-6},
	number = {1},
	urldate = {2026-01-28},
	journal = {Eur. Phys. J. B},
	author = {Patriarca, M. and Heinsalu, E. and Chakraborti, A.},
	month = jan,
	year = {2010},
	pages = {145--153},
}

@article{karsai_small_2011,
	title = {Small but slow world: {How} network topology and burstiness slow down spreading},
	volume = {83},
	shorttitle = {Small but slow world},
	url = {https://link.aps.org/doi/10.1103/PhysRevE.83.025102},
	doi = {10.1103/PhysRevE.83.025102},
	number = {2},
	urldate = {2026-01-19},
	journal = {Phys. Rev. E},
	author = {Karsai, M. and Kivelä, M. and Pan, R. K. and Kaski, K. and Kertész, J. and Barabási, A.-L. and Saramäki, J.},
	month = feb,
	year = {2011},
	pages = {025102},
}

@article{gallegati_worrying_2006,
	series = {Econophysics {Colloquium}},
	title = {Worrying trends in econophysics},
	volume = {370},
	issn = {0378-4371},
	url = {https://www.sciencedirect.com/science/article/pii/S0378437106004420},
	doi = {10.1016/j.physa.2006.04.029},
	number = {1},
	urldate = {2026-01-28},
	journal = {Physica A: Statistical Mechanics and its Applications},
	author = {Gallegati, Mauro and Keen, Steve and Lux, Thomas and Ormerod, Paul},
	month = oct,
	year = {2006},
	pages = {1--6},
}

@article{greenberg_twenty-five_2024,
	title = {Twenty-five years of random asset exchange modeling},
	volume = {97},
	issn = {1434-6036},
	url = {https://doi.org/10.1140/epjb/s10051-024-00695-3},
	doi = {10.1140/epjb/s10051-024-00695-3},
	number = {6},
	urldate = {2026-01-28},
	journal = {Eur. Phys. J. B},
	author = {Greenberg, Max and Gao, H. Oliver},
	month = jun,
	year = {2024},
	pages = {69},
}

@article{chakrabarti_microeconomics_2009,
	title = {Microeconomics of the ideal gas like market models},
	volume = {388},
	issn = {0378-4371},
	url = {https://www.sciencedirect.com/science/article/pii/S0378437109004865},
	doi = {10.1016/j.physa.2009.06.038},
	number = {19},
	urldate = {2026-01-28},
	journal = {Physica A: Statistical Mechanics and its Applications},
	author = {Chakrabarti, Anindya S. and Chakrabarti, Bikas K.},
	month = oct,
	year = {2009},
	pages = {4151--4158},
}

@article{mattsson_sarafu_2022,
	title = {Sarafu {Community} {Inclusion} {Currency} 2020–2021},
	volume = {9},
	copyright = {2022 The Author(s)},
	issn = {2052-4463},
	url = {https://www.nature.com/articles/s41597-022-01539-4},
	doi = {10.1038/s41597-022-01539-4},
	number = {1},
	urldate = {2026-01-28},
	journal = {Sci Data},
	author = {Mattsson, Carolina E. S. and Criscione, Teodoro and Ruddick, William O.},
	month = jul,
	year = {2022},
	pages = {426},
}

@misc{ruddick_sarafu_2021,
	type = {Data {Collection}},
	title = {Sarafu {Community} {Inclusion} {Currency}, 2020-2021},
	copyright = {ukda\_eul},
	url = {https://reshare.ukdataservice.ac.uk/855142/},
	doi = {10.5255/UKDA-SN-855142},
	urldate = {2021-11-10},
	publisher = {UK Data Service},
	author = {Ruddick, William O.},
	collaborator = {Criscione, Teodoro and Mattsson, C. E. S.},
	month = aug,
	year = {2021},
}

@inproceedings{starnini_smurf-based_2021,
	address = {Cham},
	title = {Smurf-{Based} {Anti}-money {Laundering} in {Time}-{Evolving} {Transaction} {Networks}},
	isbn = {978-3-030-86514-6},
	doi = {10.1007/978-3-030-86514-6_11},
	booktitle = {Machine {Learning} and {Knowledge} {Discovery} in {Databases}. {Applied} {Data} {Science} {Track}},
	publisher = {Springer International Publishing},
	author = {Starnini, Michele and Tsourakakis, Charalampos E. and Zamanipour, Maryam and Panisson, André and Allasia, Walter and Fornasiero, Marco and Puma, Laura Li and Ricci, Valeria and Ronchiadin, Silvia and Ugrinoska, Angela and Varetto, Marco and Moncalvo, Dario},
	editor = {Dong, Yuxiao and Kourtellis, Nicolas and Hammer, Barbara and Lozano, Jose A.},
	year = {2021},
	pages = {171--186},
}

@article{fujiwara_money_2021,
	title = {Money flow network among firms’ accounts in a regional bank of {Japan}},
	volume = {10},
	issn = {2193-1127},
	url = {https://doi.org/10.1140/epjds/s13688-021-00274-x},
	doi = {10.1140/epjds/s13688-021-00274-x},
	number = {1},
	urldate = {2026-01-29},
	journal = {EPJ Data Sci.},
	author = {Fujiwara, Yoshi and Inoue, Hiroyasu and Yamaguchi, Takayuki and Aoyama, Hideaki and Tanaka, Takuma and Kikuchi, Kentaro},
	month = apr,
	year = {2021},
	pages = {19},
}

@article{ialongo_reconstructing_2022,
	title = {Reconstructing firm-level interactions in the {Dutch} input–output network from production constraints},
	volume = {12},
	copyright = {2022 The Author(s)},
	issn = {2045-2322},
	url = {https://www.nature.com/articles/s41598-022-13996-3},
	doi = {10.1038/s41598-022-13996-3},
	number = {1},
	urldate = {2026-01-29},
	journal = {Sci Rep},
	author = {Ialongo, Leonardo Niccolò and de Valk, Camille and Marchese, Emiliano and Jansen, Fabian and Zmarrou, Hicham and Squartini, Tiziano and Garlaschelli, Diego},
	month = jul,
	year = {2022},
	pages = {11847},
}

@misc{bacilieri_firm-level_2023,
	address = {Oxford},
	title = {Firm-level production networks: what do we (really) know?},
    url = {https://www.inet.ox.ac.uk/publications/no-2025-14-firm-level-production-networks-what-do-we-really-know},
	publisher = {INET Oxford Working Paper},
	author = {Bacilieri, Andrea and Borsos, Andras and Astudillo-Estevez, Pablo and Lafond, Francois},
	month = may,
	year = {2023},
}

@article{letizia_corporate_2019,
	title = {Corporate payments networks and credit risk rating},
	volume = {8},
	issn = {2193-1127},
	url = {https://doi.org/10.1140/epjds/s13688-019-0197-5},
	doi = {10.1140/epjds/s13688-019-0197-5},
	number = {1},
	urldate = {2026-01-29},
	journal = {EPJ Data Sci.},
	author = {Letizia, Elisa and Lillo, Fabrizio},
	month = jun,
	year = {2019},
	pages = {21},
}

@article{meiklejohn_fistful_2016,
	title = {A fistful of {Bitcoins}: characterizing payments among men with no names},
	volume = {59},
	issn = {0001-0782},
	shorttitle = {A fistful of {Bitcoins}},
	url = {https://dl.acm.org/doi/10.1145/2896384},
	doi = {10.1145/2896384},
	number = {4},
	urldate = {2026-01-29},
	journal = {Commun. ACM},
	author = {Meiklejohn, Sarah and Pomarole, Marjori and Jordan, Grant and Levchenko, Kirill and McCoy, Damon and Voelker, Geoffrey M. and Savage, Stefan},
	month = mar,
	year = {2016},
	pages = {86--93},
}

@misc{coquide_orbitaal_2024,
	title = {{ORBITAAL}: {A} {Temporal} {Graph} {Dataset} of {Bitcoin} {Entity}-{Entity} {Transactions}},
	shorttitle = {{ORBITAAL}},
	url = {http://arxiv.org/abs/2408.14147},
	doi = {10.48550/arXiv.2408.14147},
	urldate = {2024-09-05},
	publisher = {arXiv},
	author = {Coquidé, Célestin and Cazabet, Rémy},
	month = aug,
	year = {2024},
	note = {arXiv:2408.14147 [physics]},
}

@article{collibus_microvelocity_2025,
	title = {The microvelocity of money in {Ethereum}},
	volume = {14},
	issn = {2193-1127},
	url = {https://doi.org/10.1140/epjds/s13688-024-00518-6},
	doi = {10.1140/epjds/s13688-024-00518-6},
	number = {1},
	urldate = {2026-01-29},
	journal = {EPJ Data Sci.},
	author = {De Collibus, Francesco Maria and Campajola, Carlo and Tessone, Claudio J.},
	month = feb,
	year = {2025},
	pages = {11},
}

@article{chatterjee_kinetic_2007,
	title = {Kinetic exchange models for income and wealth distributions},
	volume = {60},
	issn = {1434-6036},
	url = {https://doi.org/10.1140/epjb/e2007-00343-8},
	doi = {10.1140/epjb/e2007-00343-8},
	number = {2},
	urldate = {2026-01-29},
	journal = {Eur. Phys. J. B},
	author = {Chatterjee, A. and Chakrabarti, B. K.},
	month = nov,
	year = {2007},
	pages = {135--149},
}

@article{campajola_microvelocity_2022,
	title = {{MicroVelocity}: rethinking the {Velocity} of {Money} for digital currencies},
	shorttitle = {{MicroVelocity}},
	url = {http://arxiv.org/abs/2201.13416},
	urldate = {2022-02-11},
	journal = {arXiv:2201.13416 [physics, q-fin]},
	author = {Campajola, Carlo and D'Errico, Marco and Tessone, Claudio J.},
	month = jan,
	year = {2022},
	note = {arXiv: 2201.13416},
}

@book{mcneil_quantitative_2015,
	title = {Quantitative {Risk} {Management}: {Concepts}, {Techniques} and {Tools} - {Revised} {Edition}},
	isbn = {978-0-691-16627-8},
	shorttitle = {Quantitative {Risk} {Management}},
	publisher = {Princeton University Press},
	author = {McNeil, Alexander J. and Frey, Rüdiger and Embrechts, Paul},
	month = may,
	year = {2015},
}

@article{semeraro_structural_2020,
	title = {Structural inequalities emerging from a large wire transfers network},
	volume = {5},
	issn = {2364-8228},
	url = {https://doi.org/10.1007/s41109-020-00314-x},
	doi = {10.1007/s41109-020-00314-x},
	number = {1},
	urldate = {2026-02-02},
	journal = {Appl Netw Sci},
	author = {Semeraro, Alfonso and Tambuscio, Marcella and Ronchiadin, Silvia and Li Puma, Laura and Ruffo, Giancarlo},
	month = oct,
	year = {2020},
	pages = {76},
}

@article{arthur_foundations_2021,
	title = {Foundations of complexity economics},
	volume = {3},
	copyright = {2021 Springer Nature Limited},
	issn = {2522-5820},
	url = {https://www.nature.com/articles/s42254-020-00273-3},
	doi = {10.1038/s42254-020-00273-3},
	number = {2},
	urldate = {2026-02-02},
	journal = {Nat Rev Phys},
	author = {Arthur, W. Brian},
	month = feb,
	year = {2021},
	pages = {136--145},
}

@article{boss_network_2004,
	title = {Network topology of the interbank market},
	volume = {4},
	issn = {1469-7688},
	url = {https://www.tandfonline.com/doi/abs/10.1080/14697680400020325},
	doi = {10.1080/14697680400020325},
	number = {6},
	urldate = {2026-02-02},
	journal = {Quantitative Finance},
	author = {Boss, Michael and Elsinger, Helmut and Summer, Martin and Thurner 4, Stefan},
	month = dec,
	year = {2004},
	pages = {677--684},
}

@article{mcnerney_how_2022,
	title = {How production networks amplify economic growth},
	volume = {119},
	url = {https://www.pnas.org/doi/full/10.1073/pnas.2106031118},
	doi = {10.1073/pnas.2106031118},
	number = {1},
	urldate = {2026-02-02},
	journal = {Proceedings of the National Academy of Sciences},
	author = {McNerney, James and Savoie, Charles and Caravelli, Francesco and Carvalho, Vasco M. and Farmer, J. Doyne},
	month = jan,
	year = {2022},
	pages = {e2106031118},
}

@article{diem_quantifying_2022,
	title = {Quantifying firm-level economic systemic risk from nation-wide supply networks},
	volume = {12},
	copyright = {2022 The Author(s)},
	issn = {2045-2322},
	url = {https://www.nature.com/articles/s41598-022-11522-z},
	doi = {10.1038/s41598-022-11522-z},
	number = {1},
	urldate = {2026-02-02},
	journal = {Sci Rep},
	author = {Diem, Christian and Borsos, András and Reisch, Tobias and Kertész, János and Thurner, Stefan},
	month = may,
	year = {2022},
	pages = {7719},
}

@article{battiston_debtrank_2012,
	title = {{DebtRank}: {Too} {Central} to {Fail}? {Financial} {Networks}, the {FED} and {Systemic} {Risk}},
	volume = {2},
	copyright = {2012 The Author(s)},
	issn = {2045-2322},
	shorttitle = {{DebtRank}},
	url = {https://www.nature.com/articles/srep00541},
	doi = {10.1038/srep00541},
	number = {1},
	urldate = {2026-02-02},
	journal = {Sci Rep},
	author = {Battiston, Stefano and Puliga, Michelangelo and Kaushik, Rahul and Tasca, Paolo and Caldarelli, Guido},
	month = aug,
	year = {2012},
	pages = {541},
}

@article{mandelbrot_pareto-levy_1960,
	title = {The {Pareto}-{Lévy} {Law} and the {Distribution} of {Income}},
	volume = {1},
	issn = {0020-6598},
	url = {https://www.jstor.org/stable/2525289},
	doi = {10.2307/2525289},
	number = {2},
	urldate = {2026-02-02},
	journal = {International Economic Review},
	author = {Mandelbrot, Benoit},
	year = {1960},
	pages = {79--106},
}

@article{axtell_zipf_2001,
	title = {Zipf {Distribution} of {U}.{S}. {Firm} {Sizes}},
	volume = {293},
	url = {https://www.science.org/doi/10.1126/science.1062081},
	doi = {10.1126/science.1062081},
	number = {5536},
	urldate = {2026-02-02},
	journal = {Science},
	author = {Axtell, Robert L.},
	month = sep,
	year = {2001},
	pages = {1818--1820},
}

@article{hotte_mapping_2025,
	title = {Mapping the disaggregated economy in real-time: using granular payment network data to complement national accounts},
	volume = {37},
	issn = {0953-5314},
	shorttitle = {Mapping the disaggregated economy in real-time},
	url = {https://doi.org/10.1080/09535314.2025.2518944},
	doi = {10.1080/09535314.2025.2518944},
	number = {3},
	urldate = {2026-02-03},
	journal = {Economic Systems Research},
	author = {Hötte, Kerstin},
	month = jul,
	year = {2025},
	pages = {427--454},
}

@article{mattsson_network_2025,
	title = {Network growth under opportunistic attachment},
	volume = {10},
	issn = {2364-8228},
	url = {https://doi.org/10.1007/s41109-025-00706-x},
	doi = {10.1007/s41109-025-00706-x},
	language = {en},
	number = {1},
	urldate = {2026-02-10},
	journal = {Appl Netw Sci},
	author = {Mattsson, Carolina E. S.},
	month = jun,
	year = {2025},
	pages = {21},
}








\end{document}